\documentclass[a4paper,10pt]{article}
\usepackage{graphicx}
\usepackage{comment}
\usepackage{amssymb}
\usepackage{amsmath}
\usepackage{nicefrac}
\usepackage{graphicx}
\usepackage{dcolumn}
\usepackage{bm}
\usepackage{bbm}
\usepackage{comment}
\usepackage{multirow}
\usepackage{cite}
\usepackage{mathrsfs}
\usepackage{url}

\newcommand{\beq}{\begin{equation}}
\newcommand{\eeq}{\end{equation}}
\newcommand{\beqa}{\begin{eqnarray}}
\newcommand{\eeqa}{\end{eqnarray}}
\newcommand{\hA}{\hat{A}}
\newcommand{\hB}{\hat{B}}
\newcommand{\ha}{\hat{a}}

\begin{document}

\huge

\begin{center}
Closed forms of the Zassenhaus formula
\end{center}

\vspace{0.5cm}

\large

\begin{center}
L\'eonce Dupays$^{a,b,}$\footnote{leonce.dupays@uni.lu} and Jean-Christophe Pain$^{c,d}$
\end{center}

\normalsize

\begin{center}
\it $^a$Department of Physics and Materials Science, University of Luxembourg, L-1511 Luxembourg, G. D. Luxembourg\\
\it $^b$Donostia International Physics Center, E-20018 San Sebasti\'an, Spain\\
\it $^c$CEA, DAM, DIF, F-91297 Arpajon, France\\
\it $^d$Universit\'e Paris-Saclay, CEA, Laboratoire Mati\`ere en Conditions Extr\^emes,\\
\it 91680 Bruy\`eres-le-Ch\^atel, France
\end{center}

\abstract{The Zassenhaus formula finds many applications in theoretical physics or mathematics, from fluid dynamics to differential geometry. The non-commutativity of the elements of the algebra implies that the exponential of a sum of operators cannot be expressed as the product of exponentials of operators. The exponential of the sum can then be decomposed as the product of the exponentials multiplied by a supplementary term which takes generally the form of an infinite product of exponentials. Such a procedure is often referred to as ``disentanglement''. However, for some special commutators, closed forms can be found. In this work, we propose a closed form for the Zassenhaus formula when the commutator of operators $\hat{X}$ and $\hat{Y}$ satisfy the relation $[\hat{X},\hat{Y}]=u\hat{X}+v\hat{Y}+c\mathbbm{1}$. Such an expression boils down to three equivalent versions, a left-sided, a centered and a right-sided formula:
\begin{equation*}
e^{\hat{X}+\hat{Y}}=e^{\hat{X}}e^{\hat{Y}}e^{g_{r}(u,v)[\hat{X},\hat{Y}]}=e^{\hat{X}}e^{g_{c}(u,v)[\hat{X},\hat{Y}]}e^{\hat{Y}}=e^{g_{\ell}(u,v)[\hat{X},\hat{Y}]}e^{\hat{X}}e^{\hat{Y}},
\end{equation*}
with respective arguments,
\begin{eqnarray*}
g_{r}(u,v)&=&g_{c}(v,u)e^{u}=g_{\ell}(v,u)=\frac{u\left(e^{u-v}-e^{u}\right)+v\left(e^{u}-1\right)}{vu(u-v)}
\end{eqnarray*}
for $u\ne v$ and 
\begin{eqnarray*}
g_{r}(u,u)=\frac{u+1-e^u}{u^2}\;\;\;\;\mathrm{with}\;\;\;\; g_r(0,0)=-1/2.
\end{eqnarray*}
With additional special case
\begin{eqnarray*} 
g_{r}(0,v)= -\frac{e^{-v}-1+v}{v^{2}}, \quad & g_{r}(u,0)=\frac{e^{u}(1-u)-1}{u^{2}}.
\end{eqnarray*}
}

\section{Introduction}\label{sec1}

The Zassenhaus formula \cite{Kimura_2017,Wang_2019} plays an important role in various fields of physics, such as the Dirac monopole problem \cite{Soloviev_2016}, quantum spin lattices \cite{Braz_2016}, fluids dynamics \cite{Geiser_2013} and the study of solitary waves \cite{Tumer2017}, statistical mechanics \cite{Sornborger_1999}, many-body theories or quantum optics. In particle accelerator physics, the Zassenhaus formula was successfully used to compute the relevant maps both in Taylor-series and factorized-product forms \cite{Dragt_2009}. 
It also worth mentioning that a new family of high-order splitting methods for the numerical integration of the time-dependent Schr\"odinger equation based on a symmetric version of the Zassenhaus formula was found \cite{Iserles_2011,Bader2016}. The Zassenhaus formula is also of fundamental mathematical interests \cite{Bayen_1979,Suzuki_1977,Katriel_1996,Scholz_2006,Ebrahimi_2007,Arnal_2017}, for instance in order to disentangle exponential operators \cite{sridhar_2002,Quesne_2004} or in differential geometry \cite{Nishimura_2013}. This formula is the dual formula of the well-known Baker-Campbell-Hausdorff\ (BCH) formula which aims at composing two operator exponentials. When two operators $\hat{X}$ and $\hat{Y}$ 
do not commute, expanding $e^{\hat{X}+\hat{Y}}$ is a cumbersome task. In the context of quantum dynamics, some approximate forms have been proposed, such as the Trotter-Suzuki decomposition \cite{Suzuki_1985} or the Magnus expansion \cite{Magnus1954}. Recently, new methods were suggested, for instance using recurrence relations \cite{Casas_2012}. Nevertheless, it appears that for some special commutators, the expansion reduces to a closed form, the most famous one being the Glauber formula
\begin{equation}\label{glauber}
e^{\hat{X}+\hat{Y}}=e^{\hat{X}}e^{\hat{Y}}e^{-\frac{1}{2}[\hat{X},\hat{Y}]}
\end{equation}
for $[\hat{X},[\hat{X},\hat{Y}]]=0$ and $[\hat{Y},[\hat{X},\hat{Y}]]=0$. However, if many articles are devoted to the search for closed forms of the BCH formula \cite{Van_Brunt_2015,Matone_2016,Matone_2015,Van_Brunt_2018,Pain_2012} and if some exotic forms of BCH have been found \cite{Mostovoy_2016}, it seems that the Zassenhaus formula \cite{Magnus_1954} has been less investigated. In the present work, we show that a closed form of the Zassenhaus formula can be derived when $\hat{X}$ and $\hat{Y}$ satisfy the peculiar commutation relation $[\hat{X},\hat{Y}]=u\hat{X}+v\hat{Y}+c\mathbbm{1}$. More precisely, such a formula covers three forms, a left-sided, a centered and a right-sided one:
\begin{equation}
e^{\hat{X}+\hat{Y}}=e^{\hat{X}}e^{\hat{Y}}e^{g_{r}(u,v)[\hat{X},\hat{Y}]}=e^{\hat{X}}e^{g_{c}(u,v)[\hat{X},\hat{Y}]}e^{\hat{Y}}=e^{g_{\ell}(u,v)[\hat{X},\hat{Y}]}e^{\hat{X}}e^{\hat{Y}},
\end{equation}
with respective arguments
\begin{equation}
g_{r}(u,v)=g_{c}(v,u)e^{u}=g_{\ell}(v,u)=\frac{u\left(e^{u-v}-e^{u}\right)+v\left(e^{u}-1\right)}{vu(u-v)},
\end{equation}
for $u\ne v$ and 
\begin{eqnarray}
g_{r}(u,u)=\frac{u+1-e^u}{u^2},
\end{eqnarray}
with $g_r(0,0)=-1/2$. The important subcases $v=c=0$ and $u=c=0$ have been already treated in \cite{Suzuki_1985}:
\begin{equation} 
g_{r}(0,v)= -\frac{e^{-v}-1+v}{v^{2}}, 
\end{equation}
and
\begin{equation}
g_{r}(u,0)=\frac{e^{u}(1-u)-1}{u^{2}}.
\end{equation}

In section \ref{sec2}, we derive our main results with a careful treatment of the special cases. We provide three different proofs of that expressions; the first one relies on an integral representation published by Suzuki \cite{Suzuki_1985}. The second one is based on the BCH formula, using the closed form published by Van Brunt and Visser \cite{Van_Brunt_2015}. The third one consists in using the recursion relation developed by Casas \emph{et al.} \cite{Casas_2012}. Consequences of the new relation and examples of applications are given in section \ref{sec3}. A centered form of the relation is obtained. Possible applications to specific Lie algebras such as $SU(1,1)$ or the algebra of Linbladian operator of a qubit under dissipation are outlined. 
\section{A closed form of the Zassenhaus formula}\label{sec2}
\subsection{First proof involving Suzuki's integral relation}\label{subsec21}

The Zassenhaus theorem states that for $\hat{X},\hat{Y}$ generators of a Lie algebra, $e^{\hat{X}+\hat{Y}}$ can be uniquely decomposed in the following way: 
\begin{equation}
\label{eq:Zass}
e^{\hat{X}+\hat{Y}}=e^{\hat{X}}e^{\hat{Y}}\prod_{n=2}^{\infty}e^{C_{n}(\hat{X},\hat{Y})}=e^{\hat{X}}e^{\hat{Y}}e^{C_{2}(\hat{X},\hat{Y})}e^{C_{3}(\hat{X},\hat{Y})}\dots e^{C_{n}(\hat{X},\hat{Y})}\dots,
\end{equation}
where $C_{n}(\hat{X},\hat{Y})$ is a Lie polynomial in $\hat{X}$ and $\hat{Y}$ of degree $n$.
The first terms are
\begin{eqnarray}
e^{\hat{X}+\hat{Y}}&=&e^{\hat{X}}e^{\hat{Y}}e^{-\frac{1}{2}[\hat{X},\hat{Y}]}e^{\frac{1}{6}\left(2[\hat{Y},[\hat{X},\hat{Y}]]+[\hat{X},[\hat{X},\hat{Y}]]\right)}\nonumber\\
& &\times e^{-\frac{1}{24}\left([[[\hat{X},\hat{Y}],\hat{X}],\hat{X}]+3[[[\hat{X},\hat{Y}],\hat{X}],\hat{Y}]+3[[[\hat{X},\hat{Y}],\hat{Y}],\hat{Y}]\right)}\dots
\end{eqnarray}
In the case of the specific commutation relation $[\hat{X},\hat{Y}]=u\hat{X}+v\hat{Y}+c\mathbbm{1}$, the multiple commutators in the argument of the exponentials all reduce to a constant multiplied by $[\hat{X},\hat{Y}]$. Equation (\ref{eq:Zass}) may be therefore recast in the form $e^{\hat{X}+\hat{Y}}=e^{\hat{X}}e^{\hat{Y}}e^{\mu[\hat{X},\hat{Y}]}$, where $\mu$ is to be determined. We denote $\mathrm{ad}_{\hA}$ the adjoint operator of an operator $\hA$, defined by recurrence as $\mathrm{ad}^{n}_{\hA}\hB=[\hA,\mathrm{ad}^{n-1}_{\hA}\hB]$ with $\mathrm{ad}^{0}_{\hA}\hB=\hB$ so that it is possible to rewrite $e^{\hA}\hB e^{-\hA}=e^{\mathrm{ad}_{\hA}}\hB$. A first method to determine this constant consists in an integral representation of the Zassenhaus formula, introducing a new variable $t$ in the exponential \cite{Suzuki_1985}
\begin{equation}
\label{eq:integration_zassenhaus}
e^{t\left(\hat{X}+\hat{Y}\right)}=e^{t\hat{X}}e^{t\hat{Y}}\mathcal{T}\exp_{+}\left(\int_{0}^{t}dse^{-s~\mathrm{ad}_{\hat{Y}}}\left(e^{-s~\mathrm{ad}_{\hat{X}}}-1\right)\hat{Y}\right),
\end{equation}
where $\mathcal{T}$ and $+$ stand for normal ordering. The latter formula is derived in \ref{appA}. The time-ordered operator $U(t,t_{0})=\mathcal{T}\exp_{+}\left(\int_{t_{0}}^{t}H(s)ds\right)$ can be explicitly written
\begin{equation}
U(t,t_{0})=\mathbbm{1}+\sum_{n=1}^{\infty}\frac{1}{n!}\int_{t_{0}}^{t}dt_{1}\int_{t_0}^{t_1}dt_2\cdots\int_{t_0}^{t_{n-1}}dt_n ~H(t_1)H(t_2)\cdots H(t_n). \label{eq:time_ordered}
\end{equation}
This unitary evolution can be written in function of the normal ordering operator $\mathcal{T}$ defined as 
\begin{equation}
\mathcal{T}[H(t_1)H(t_2)\cdots H(t_n)]=H(t_{i_1})H(t_{i_2})\cdots H(t_{i_n}),
\end{equation}
with
\begin{equation}
t_{i_1}>t_{i_2}>\cdots>t_{i_{n}},
\end{equation}
simplifying (\ref{eq:time_ordered}) in \cite{Huang_2010}
\begin{equation}
U(t,t_{0})=\mathbbm{1}+\sum_{n=1}^{\infty}\frac{1}{n!}\int_{t_{0}}^{t}dt_{1}\int_{t_0}^{t}dt_2\cdots\int_{t_0}^{t}dt_n~\mathcal{T}[H(t_1)H(t_2)\cdots H(t_n)].
\end{equation}
Moreover for the special commutator $[\hat{X},\hat{Y}]=u\hat{X}+v\hat{Y}+c\mathbbm{1}$, one has 
\begin{equation}
\label{eq:analytical}
H(s)=e^{-s~\mathrm{ad}_{\hat{Y}}}\left(e^{-s~\mathrm{ad}_{\hat{X}}}-1\right)\hat{Y}=\frac{e^{s(u-v)}-e^{su}}{v}[\hat{X},\hat{Y}].
\end{equation}
One can note that the dependence in the parameter $c$ in the previous expression only appears in the commutator $[\hat{X},\hat{Y}]$ due to the fact that the adjoint operator gives a null result when it acts on the identity operator. The specific form of $H(s)$ guarantees commutation relations for the Hamiltonian at different times $[H(t_i),H(t_j)]=0$ for $t_i\neq t_j$, which further simplify the previous expression
\begin{eqnarray}
U(t,t_{0})&=&\mathbbm{1}+\sum_{n=1}^{\infty}\frac{1}{n!}\left(\int_{t_{0}}^{t}dt_{1}H(t_1)\right)^{n}=\exp\left[\int_{t_{0}}^{t}dt_{1}H(t_1)\right].
\end{eqnarray}
Inserting Eq. (\ref{eq:analytical}) into Eq. (\ref{eq:integration_zassenhaus}) allows one to perform the integration. Then, choosing $t=1$ gives the closed-form expression for the Zassenhaus formula
\begin{equation}
\label{eq:main_result}
e^{\hat{X}+\hat{Y}}=e^{\hat{X}}e^{\hat{Y}}\exp\left(g_{r}(u,v)[\hat{X},\hat{Y}]\right),
\end{equation}
with the right-sided coefficient 
\begin{equation}
\label{eq:main_result2}
g_{r}(u,v)=\frac{u\left(e^{u-v}-e^{u}\right)+v\left(e^{u}-1\right)}{uv(u-v)}.
\end{equation}
Equations (\ref{eq:main_result}) and (\ref{eq:main_result2}) constitute the main results of the present work.
\subsection{Special cases}\label{subsec24}
In order to compute $g_{r}(u,v)$ in the special cases $u=0$, $v=0$ or $u=v=0$, De l'H\^opital's rule yields
\begin{equation}
g_{r}(0,v)= -\frac{e^{-v}-1+v}{v^{2}}, 
\end{equation}
and
\begin{equation}
g_{r}(u,0)=\frac{e^{u}(1-u)-1}{u^{2}},
\end{equation}
as well as
\begin{equation}
\quad g_{r}(0,0)=-\frac{1}{2}.
\end{equation}
For a commutator of the form $[\hat{X},\hat{Y}]=c\mathbbm{1}$, one recovers the Glauber formula \ref{glauber}. Finally, the case $u=v$ is also interesting:
\begin{equation}
g_{r}(u,u)=\frac{u+1-e^{u}}{u^{2}}
\end{equation}
and special forms of $g_{\ell}$ and $g_{c}$ are directly found from $g_{r}$.

\subsection{Second proof: starting from the BCH formula}\label{subsec22}
It is possible to derive Eqs. (\ref{eq:main_result}) and (\ref{eq:main_result2}) starting from a closed form of the BCH formula. For pedagogical purposes, we detail below the calculation of the left-sided expression of the Zassenhaus formula (with coefficient $g_{\ell}(u,v)$) for the special commutator $[\hat{X},\hat{Y}]=u\hat{X}+v\hat{Y}+c\mathbbm{1}$. Let us look for $g_{\ell}$ satisfying
\begin{equation}\label{prem}
e^{\hat{X}+\hat{Y}}=e^{g_{\ell}(u,v)[\hat{X},\hat{Y}]}e^{\hat{X}}e^{\hat{Y}}.
\end{equation}
It was demonstrated in \cite{Van_Brunt_2015} that for the special commutator $[\hat{X},\hat{Y}]=u\hat{X}+v\hat{Y}+c\mathbbm{1}$, the BCH formula reads
\begin{equation}\label{vbw}
e^{\hat{X}}e^{\hat{Y}}=e^{\hat{X}+\hat{Y}+f(u,v)[\hat{X},\hat{Y}]},
\end{equation}
with for $u\neq v$
\begin{eqnarray}
f(u,v)&=&\frac{ue^{u}(e^{v}-1)-ve^{v}(e^{u}-1)}{uv(e^{u}-e^{v})}.
\end{eqnarray}
 Eq. (\ref{prem}) becomes
\begin{equation}\label{connec}
e^{\hat{X}+\hat{Y}}=e^{g_{\ell}(u,v)[\hat{X},\hat{Y}]}e^{\hat{X}+\hat{Y}+f(u,v)[\hat{X},\hat{Y}]}=e^{\hA}e^{\hB},
\end{equation}
where we have set $\hA=g_{\ell}(u,v)[\hat{X},\hat{Y}]$ and $\hB=\hat{X}+\hat{Y}+f(u,v)[\hat{X},\hat{Y}]$. One has
\begin{eqnarray}
[\hA,\hB]&=&g_{\ell}(u,v)[[\hat{X},\hat{Y}],\hat{X}]+g_{\ell}(u,v)[[\hat{X},\hat{Y}],\hat{Y}]\nonumber\\
&=&g_{\ell}(u,v)[u\hat{X}+v\hat{Y}+c\mathbbm{1},\hat{X}]+g_{\ell}(u,v)[u\hat{X}+v\hat{Y}+c\mathbbm{1},\hat{Y}]\nonumber\\
&=&g_{\ell}(u,v)v[\hat{Y},\hat{X}]+g_{\ell}(u,v)u[\hat{X},\hat{Y}]\nonumber\\
&=&(u-v)\hA.
\end{eqnarray}
Thus, $[\hA,\hB]=\tilde{u}\hA+\tilde{v}\hB+\tilde{c}1$, with $\tilde{u}=u-v$, $\tilde{v}=0$ and $\tilde{c}=0$, and the operators $\hA$ and $\hB$ satisfy the same commutation relation as $\hat{X}$ and $\hat{Y}$. They obey therefore the special case of the BCH identity (\ref{vbw}),
\begin{equation}
e^{\hA}e^{\hB}=e^{\hA+\hB+f(u-v,0)[\hA,\hB]},
\end{equation}
and therefore Eq. (\ref{connec}) becomes
\begin{eqnarray}
e^{\hat{X}+\hat{Y}}&=&e^{\hat{X}+\hat{Y}+[g_{\ell}(u,v)+f(u,v)][\hat{X},\hat{Y}]+f(u-v,0)[\hA,\hB]}\nonumber\\
&=&e^{\hat{X}+\hat{Y}+\left\{g_{\ell}(u,v)+f(u,v)+f(u-v,0)(u-v)g_{\ell}(u,v)\right\}[\hat{X},\hat{Y}]}
\end{eqnarray}
yielding
\begin{equation}
g_{\ell}(u,v)=-\frac{f(u,v)}{1+f(u-v,0)(u-v)}.
\end{equation}
Since 
\begin{equation}
\lim_{y\rightarrow 0}f(x,y)=\frac{e^x}{e^x-1}-\frac{1}{x},
\end{equation} 
one gets
\begin{equation}
g_{\ell}(u,v)=\frac{v\left(e^{v-u}-e^v\right)+u\left(e^v-1\right)}{(v-u)uv}.
\end{equation}

\subsection{Third proof: deduction from recurrence relations}\label{subsec23}

Casas \emph{et al.} developed a technique to compute recursively the exponents of the Zassenhaus formula \cite{Casas_2012}. However, it can be shown that if $\hat{X}$ and $\hat{Y}$ verify the above mentioned commutation relation, their work enables one to recover the relation (\ref{eq:main_result2}).

\subsubsection{Recursive computation of the Zassenhaus formula}\label{subsubsubsec231}

To find the terms of the Zassenhaus formula by recurrence, Casas \emph{et al.} \cite{Casas_2012} introduce a differentiable formula with respect to $t$ that recovers the Zassenhaus formula (\ref{eq:Zass}) for $t=1$ (throughout this section we omit the ``hat'' for some functions of operators for a sake of simplicity and when there is no ambiguity):
\begin{equation}
e^{t(\hat{X}+\hat{Y})}=e^{t \hat{X}}e^{t \hat{Y}}e^{t^{2}C_{2}}e^{t^{3}C_{3}}...
\end{equation}
Considering the following compositions
\begin{equation}
R_{1}(t)=e^{-tY}e^{-t X}e^{t(X+Y)}.
\end{equation}
as well as, for each $n\geq2$,
\begin{equation}
R_{n}(t)=e^{-t^{n}C_{n}}\dots e^{-t^{2}C_{2}}e^{-t Y}e^{-t X}e^{t(X+Y)}=e^{-t^{n}C_{n}}R_{n-1}(t),
\end{equation}
the authors introduce the functions
\begin{equation}
F_{n}: t\longmapsto \left(\frac{d}{dt}R_{n}(t)\right)R_{n}(t)^{-1}
\end{equation}
that allow one to determine the polynomial $C_{n}$ through the recurrence relations 
\begin{equation}
\label{eq1}
F_{n} (t) =e^{-t^{n}~\mathrm{ad}_{C_{n}}}\left( F_{n-1}(t) -\frac{t^{n-1}}{(n-1)!} F_{n-1}^{(n-1)}(0)\right),
\end{equation}
and
\begin{equation}
\label{eq1bis}
C_{n}=\frac{1}{n!}F_{n-1}^{(n-1)}(0),
\end{equation}
where $F^{(n)}$ denotes the $n^{th}$ derivative of $F$: $F^{(n)}=\partial^{n} F/\partial t^{n}$. One has
\begin{equation}
F_{1}(t)=e^{-t~\mathrm{ad}_{\hat{Y}}}\left(e^{-t~\mathrm{ad}_{\hat{X}}}-1\right) \hat{Y},
\end{equation}
which enables one to deduce $C_{2}$ from the knowledge of $F_{1}$, $C_{3}$ from the knowledge of $F_{2}$, and so on. A recurrence scheme to determine the coefficients $C_{n}$ is then established.

\subsubsection{Impact of the commutation relation $[\hat{X},\hat{Y}]=u\hat{X}+v\hat{Y}+c\mathbbm{1}$ on the recurrence}\label{subsubsubsec232}

Actually, the latter recurrence scheme can be exploited in order to find directly the $C_{n}$ coefficients without any recurrence if the commutator between $\hat{X}$ and $\hat{Y}$ is appropriate. If $\hat{X}$ and $\hat{Y}$ are two elements of an algebra obeying $[\hat{X},\hat{Y}]=u\hat{X}+v\hat{Y}+c\mathbbm{1}$, it is then obvious that $[\hat{X},[\hat{X},\hat{Y}]]=v[\hat{X},\hat{Y}]$ and $[\hat{Y},[\hat{X},\hat{Y}]]=-u[\hat{X},\hat{Y}]$. Subsequently, $F_{1}$ can be written 
\begin{equation}\label{F1}
F_{1}(t)=e^{-t~\mathrm{ad}_{\hat{Y}}}(e^{-t~\mathrm{ad}_{\hat{X}}}-1)Y=\frac{e^{tu}(e^{-t v}-1)}{ v}[\hat{X},\hat{Y}].
\end{equation}
Let us prove by induction that $F_{n}$ is always proportional to the commutator $[\hat{X},\hat{Y}]$, so that $F_{n}=\beta_{n}(t)[\hat{X},\hat{Y}]$ where $\beta_{n}$ is a function of $t$. This is true for $F_1(t)$ (see Eq. (\ref{F1})). Let us consider 
\begin{eqnarray}
F_{n+1} (t) =e^{-t^{n+1}~ \mathrm{ad}_{C_{n+1}}}\left( F_{n}(t) -\frac{t^{n}}{n!} F_{n}^{(n)}(0)\right).
\end{eqnarray}
According to the induction assumption (see Eq. (\ref{eq1})),
\begin{equation}
C_{n+1}=\frac{1}{(n+1)!} \left( \frac{\partial^{n}}{\partial t^{n}}\beta_{n} \right) (0) [\hat{X},\hat{Y}].
\end{equation}
The adjoint operator commutes with the term on which it acts, leading to the simplification
\begin{equation}
F_{n+1}(t)=\left[\beta_{n}(t)-\frac{t^{n}}{n!}\left(\frac{\partial^{n}}{\partial t^{n}}\beta_{n}\right)(0)\right][\hat{X},\hat{Y}],\\
\end{equation}
so that, 
\begin{equation}
\label{eq2}
\beta_{n+1}(t)=\beta_{n}(t)-\frac{t^{n}}{n!}\left(\frac{\partial^{n}}{\partial t^{n}}\beta_{n}\right)(0),
\end{equation}
proving the induction. The $F_{n}$ being of the form $F_{n}=\beta_{n}[\hat{X},\hat{Y}]$, the coefficients $C_{n}$ can be deduced directly.

\subsubsection{Expression of the Zassenhaus formula in closed form}\label{subsubsec233}

It is now possible to find a closed form of the Zassenhaus formula by summing up all the $C_{n}$. The coefficient $C_{n+1}$ reads 
\begin{equation}
C_{n+1}=\frac{1}{(n+1)!} \left( \frac{\partial^{n} }{\partial t^{n}}\beta_{n} \right)(0) [\hat{X},\hat{Y}],
\end{equation}
$ \forall~ n \geq 1$. According to Eq. (\ref{eq2}), one has
\begin{equation}
\frac{\partial^{n+1}}{\partial t^{n+1}} \beta_{n+1}=\frac{\partial^{n+1}}{\partial t^{n+1}} \beta_{n},
\end{equation}
from which we get, $ \forall~ n \geq 1$, 
\begin{eqnarray}
C_{n+1}&=&\frac{1}{(n+1)!} \left( \frac{\partial^{n} }{\partial t^{n}}\beta_{n} \right)(0) [\hat{X},\hat{Y}]=\frac{1}{(n+1)!} \left( \frac{\partial^{n} }{\partial t^{n}}\beta_{1} \right)(0) [\hat{X},\hat{Y}]\nonumber\\
&=&\frac{1}{(n+1)!} \left( \frac{\partial^{n} }{\partial t^{n}}\frac{e^{t u}(e^{-t v}-1)}{ v} \right)(0) [\hat{X},\hat{Y}].
\end{eqnarray}
Using a Cauchy product, the quantity $e^{t u}(e^{-t v}-1)/v$ can be expanded as the series
\begin{eqnarray}
\frac{e^{t u}(e^{-t v}-1)}{ v}&=&\frac{1}{v}\sum_{k=0}^{\infty}\sum_{s=0}^{k}\left(\frac{u^{s}}{s!}\frac{(-v)^{k-s}}{(k-s)!}\right)t^{k}-\frac{1}{v}\sum_{k=0}^{\infty}\frac{u^{k}}{k!}t^{k}.
\end{eqnarray}
Hence,
\begin{equation}
\frac{\partial^{n}}{\partial t^{n}}\left[\frac{e^{t u}(e^{-t v}-1)}{ v}\right]=\sum_{k=n}^{\infty}\frac{k!}{(k-n)!}\left[\frac{1}{v}\left(\sum_{s=0}^{k}\left(\frac{u^{s}}{s!}\frac{(-v)^{k-s}}{(k-s)!}\right)-\frac{u^{k}}{k!}\right)\right]t^{k-n}
\end{equation}
and
\begin{eqnarray}
\frac{\partial^{n}}{\partial t^{n}}\left[\frac{e^{t u}(e^{-t v}-1)}{ v}\right](0)&=&n!\left\{\frac{1}{v}\left(\sum_{s=0}^{n}\left(\frac{u^{s}}{s!}\frac{(-v)^{n-s}}{(n-s)!}\right)-\frac{u^{n}}{n!}\right)\right\},\nonumber\\
&=&\frac{1}{v}\left[\left(u-v \right)^{n}-u^{n}\right],
\end{eqnarray}
leading, $\forall~ n \geq 1$, to 
\begin{equation}
C_{n+1}=\frac{1}{(n+1)!}\frac{1}{v}\left[\left(u-v \right)^{n}-u^{n}\right].
\end{equation}
Finally, one obtains 
\begin{eqnarray}
g_{r}(u,v)&=&\sum_{n=2}^{\infty}C_{n}=\sum_{n=1}^{\infty}C_{n+1}=\sum_{n=1}^{\infty}\frac{1}{(n+1)!}\frac{1}{v}\left[\left(u-v \right)^{n}-u^{n}\right]\nonumber\\
&=&\frac{1}{v(u-v)}\sum_{n=1}^{\infty}\frac{(u-v)^{n+1}}{(n+1)!}-\frac{1}{vu}\sum_{n=1}^{\infty}\frac{u^{n+1}}{(n+1)!}\nonumber\\
&=&\frac{1}{v(u-v)}\sum_{n=2}^{\infty}\frac{(u-v)^{n}}{n!}-\frac{1}{vu}\sum_{n=2}^{\infty}\frac{u^{n}}{n!}\nonumber\\
&=&\frac{1}{v(u-v)}\left[e^{u-v}-1-(u-v)\right]-\frac{1}{vu}\left(e^{u}-1-u\right),
\end{eqnarray}
recovering the closed form of the Zassenhaus formula given by Eqs. (\ref{eq:main_result}) and (\ref{eq:main_result2}).

\section{Consequences and properties}\label{sec3}

\subsection{Commutation of the generators of the algebra}\label{subsec31}
If $\hat{X},\hat{Y}$ are elements of an algebra which verify the commutation relation $[\hat{X},\hat{Y}]=u\hat{X}+v\hat{Y}+c\mathbbm{1}$, we can find the commutation relation between the generators of the algebra $e^{\hat{X}}$ and $e^{\hat{Y}}$, and look for an expression of the form $e^{\hat{X}}e^{\hat{Y}}=e^{\hat{Y}}e^{\hat{X}}e^{\gamma[\hat{X},\hat{Y}]}$ where $\gamma$ is to be determined. As $\hat{X}+\hat{Y}=\hat{Y}+\hat{X}$ one can apply Eq. (\ref{eq:main_result}) for both commutators $[\hat{X},\hat{Y}]=u \hat{X}+v \hat{Y}+c\mathbbm{1}$ and $[\hat{Y},\hat{X}]=\alpha \hat{Y} +\beta \hat{X}+\delta \mathbbm{1} =-u\hat{X}-v\hat{Y}-c\mathbbm{1}$ so that
\begin{equation}
e^{\hat{X}+\hat{Y}}=e^{\hat{X}}e^{\hat{Y}}e^{g_{r}(u,v)[\hat{X},\hat{Y}]},
\end{equation}
and
\begin{equation}
e^{\hat{Y}+\hat{X}}=e^{\hat{Y}}e^{\hat{X}}e^{g_{r}(\alpha,\beta)[\hat{Y},\hat{X}]}=e^{\hat{Y}}e^{\hat{X}}e^{-g_{r}(-v,-u)[\hat{X},\hat{Y}]}.
\end{equation}
From the previous equalities, one gets 
\begin{equation}
e^{\hat{X}}e^{\hat{Y}}=e^{\hat{Y}}e^{\hat{X}}e^{-\left(g_{r}(-v,-u)+g_{r}(u,v)\right)[\hat{X},\hat{Y}]},
\label{eq:eq1}
\end{equation}
with
\begin{equation}
-g_{r}(-v,-u)-g_{r}(u,v)=\frac{1}{vu(u-v)}\{v\left(e^{-v}-1\right)\left(e^{u}-1\right)+u\left(e^{-v}-1\right)\left(1-e^{u}\right)\}.
\end{equation}
\subsection{A centered formula}\label{subsec32}

It is also possible to derive a centered formula (with coefficient $g_{c}(u,v)$) as
\begin{equation}
e^{\hat{X}+\hat{Y}}=e^{\hat{X}}e^{g_{c}(u,v)[\hat{X},\hat{Y}]}e^{\hat{Y}}.
\end{equation}
Starting from, 
\begin{equation}
\label{eq:eq2}
e^{\hat{X}+\hat{Y}}=e^{g_{\ell}(u,v)[\hat{X},\hat{Y}]}e^{\hat{X}}e^{\hat{Y}},
\end{equation}
and setting ${\hA}=g_{\ell}(u,v)[\hat{X},\hat{Y}]$ and ${\hB}=\hat{X}$, the commutator reads $[{\hA},{\hB}]=-v{\hA}$. Applying the result of Eq. (\ref{eq:eq1}), Eq. (\ref{eq:eq2}) becomes,
\begin{eqnarray}
e^{\hat{X}+\hat{Y}}&=&e^{\hB}e^{\hA}e^{-\left[g_{r}(0,v)+g_{r}(-v,0)\right][\hA,\hB]}e^{\hat{Y}}\nonumber\\
&=&e^{\hat{X}}e^{g_{\ell}(u,v)[\hat{X},\hat{Y}]}e^{v\left[g_{r}(0,v)+g_{r}(-v,0)\right]g_{\ell}(u,v)[\hat{X},\hat{Y}]}e^{\hat{Y}},
\end{eqnarray}
where
\begin{equation}
g_{c}(u,v)=\left\{1+v\left[g_{r}(0,v)+g_{r}(-v,0)\right]\right\}g_{\ell}(u,v)=e^{-v}g_{\ell}(u,v).
\end{equation}

\subsection{Exemple of applications}

It is worth mentioning that in the $SU(1,1)$ algebra $\{\ha^{\dagger2},\ha^{2},\ha^{\dagger}\ha\}$ ($\ha$ and $\ha^{\dagger}$ being annihilation and creation operators respectively), our formula can be applied to the commutators $[\ha^{\dagger2},\ha^{\dagger}\ha]=-2\ha^{\dagger2}$ and $[\ha^{2},\ha^{\dagger}\ha]=2\ha^{2}$:
\begin{eqnarray}
e^{\ha^{\dagger2}+\ha^{\dagger}\ha}&=e^{\ha^{\dagger2}}e^{\ha^{\dagger}\ha}e^{g_{r}(-2,0)[\ha^{\dagger2},\ha^{\dagger}\ha]}=e^{\ha^{\dagger2}}e^{\ha^{\dagger}\ha}e^{\frac{3e^{-2}-1}{4}\ha^{\dagger2}},\\
e^{\ha^{2}+\ha^{\dagger}\ha}&=e^{\ha^2}e^{\ha^{\dagger}\ha}e^{g_{r}(2,0)[\ha^{2},\ha^{\dagger}\ha]}=e^{\ha^{2}}e^{\ha^{\dagger}\ha}e^{-\frac{(e^{2}+1)}{4}\ha^{2}}.
\end{eqnarray}
On a different ground, the formula applies to the algebra of dissipators in open quantum systems Ref. \cite{Scopa_2019}. Consider a finite Hilbert space $\mathcal{H}_{s}$ of dimension ${\rm dim}(\mathcal{H}_{s})=n<\infty$, a vectorization procedure for operators of the form \cite{Gyamfi_2020}
\begin{eqnarray}
|\Psi\rangle \langle \phi | \rightarrow |\phi \rangle^{*} \otimes |\Psi \rangle, \quad \forall~ \Psi,\phi \in \mathcal{H}_{s}.
\end{eqnarray}
In particular, considering two operators $A$ and $B$ acting on the density matrix $\rho$, the above vectorization procedures allows one to write \cite{Gyamfi_2020}
\begin{eqnarray}
A\rho B =( B^{T}\otimes A) |\rho\rangle
\end{eqnarray}
with $|\rho\rangle$ the vectorized density matrix.
In particular, one can consider the Linbladian dissipator \cite{Rivas_2012}
\begin{equation}
\mathcal{D}_{kl}(\rho)=\sigma_{k}\rho\sigma_{l}-\frac{1}{2}\{\sigma_{l}\sigma_{k},\rho\},
\end{equation}
for a qubit, with $\sigma_{k}/\sqrt{2}$ the normalized Pauli matrices $(k=1,2,3)$, where $\{\cdot,\cdot\}$ stands for the anticommutator, which is useful to describe the dissipation of a qubit coupled to a thermal bath \cite{Rivas_2012}. The Pauli matrices being Hermitian operators, they verify $\sigma^{\dagger}_{k}=\sigma_{k}$ so that the transpose of the Pauli matrix is also equal to its complex conjugate $\sigma^{T}_{k}=(\sigma^{\dagger})^{T}=\sigma^{*}_{k}$. The dissipator can be written in the vectorized form 
\begin{eqnarray}
\mathcal{D}_{kl}&=&(\sigma_{l})^{T}\otimes \sigma_{k}-\frac{1}{2}\mathbbm{1}_{2}\otimes \sigma_{l}\sigma_{k}-\frac{1}{2}(\sigma_{k}\sigma_{l})^{T}\otimes \mathbbm{1}_{2}\nonumber\\
&=&\sigma^{*}_{l}\otimes \sigma_{k}-\frac{1}{2}\mathbbm{1}_{2}\otimes \sigma_{l}\sigma_{k}-\frac{1}{2}\sigma^{*}_{l}\sigma^{*}_{k}\otimes \mathbbm{1}_{2}.
\end{eqnarray}
Following \cite{Scopa_2019}, one can verify
\begin{eqnarray}
\mathcal{D}_{\uparrow/\downarrow}&=&\frac{1}{2}\left[\mathcal{D}_{11}+\mathcal{D}_{22}+i(\mp \mathcal{D}_{12}\pm \mathcal{D}_{21})\right],
\end{eqnarray}
and notice that 
\begin{eqnarray}
[\mathcal{D}_{\uparrow},\mathcal{D}_{\downarrow}]=\mathcal{D}_{\uparrow}-\mathcal{D}_{\downarrow}\label{eq:commutator}.
\end{eqnarray}
The closed form of the Zassenhaus formula proves useful for the previous commutator (\ref{eq:commutator}), and can be directly applied:
\begin{equation}
e^{(\alpha \mathcal{D}_{\uparrow}+\beta \mathcal{D}_{\downarrow})}=e^{\alpha\mathcal{D}_{\uparrow}}e^{\beta \mathcal{D}_{\downarrow}}e^{g_{r}(\alpha\beta,-\alpha\beta)\alpha \beta[\mathcal{D}_{\uparrow},\mathcal{D}_{\downarrow}]}=e^{\alpha\mathcal{D}_{\uparrow}}e^{\beta \mathcal{D}_{\downarrow}}e^{-\frac{(e^{\alpha \beta}-1)^{2}}{2\alpha \beta}(\mathcal{D}_{\uparrow}-\mathcal{D}_{\downarrow})}.
\end{equation}

\section{Conclusion}\label{sec4}

We obtained a new closed form of the Zassenhaus formula for a particuliar family of commutation relations. It can be given in three versions: a right-sided one, a left-sided one and a centered one. The different formulas should be useful in quantum mechanics, for instance to disentangle operators. Three different proofs were provided: the first one using a time-ordered integral relation, the second one relying on the Baker-Campbell-Hausdorff formula and the third one based on recurrence relations. Special cases were discussed, and possible applications to open quantum systems were outlined. We hope that the formalism, the derivation techniques and the results presented here will open the way to the derivation of further analytical forms of the Zassenhaus formula in specific cases and Lie algebras.

\section*{Acknowledgement} 

It is a pleasure to thank Adolfo del Campo and I\~nigo Luis Egusquiza for bright comments, as well as Yvan Castin for helpful discussions. L. Dupays would like to acknowledge DIPC for financial support at early stage of this work. 



\section*{Appendix A: Proof of relation (\ref{eq:integration_zassenhaus})}\label{appA}

One has
\begin{equation}\label{hadamard}
e^{-t~\mathrm{ad}_{\hA}}B=e^{-t\hA}~\hB~e^{t\hA}.
\end{equation}
The latter formula is well known (often referred to as Hadamard's lemma). The left-hand side reads, using the usual series expansion of the exponential
\begin{equation}
e^{-t~\mathrm{ad}_{\hA}}\hB=\hB-t[\hA,\hB]+\frac{t^2}{2}\left[\hA,\left[\hA,\hB\right]\right]+\cdots.
\end{equation}
Then, expanding the two exponentials in powers series (around $t$=0) in the right-hand side of Eq. (\ref{hadamard}) completes the proof. An alternative proof consists in noticing that
\begin{equation}
\frac{d}{dt}\left(e^{t\hA}~\hB~e^{-t\hA}\right)=\left[\hA,e^{t\hA}~\hB~e^{-t\hA}\right],
\end{equation}
and thus $f(t)=e^{t\hA}~\hB~e^{-t\hA}$ satisfies
\begin{equation}
\frac{df}{dt}=\left[\hA,f(t)\right],
\end{equation}
and iteratively
\begin{equation}
\frac{d^2f}{dt^2}=\left[\hA,\left[\hA,f(t)\right]\right],
\end{equation}
etc. Expanding $f(t)$ in power series in $t$ yields the final result.

Let us now transform the argument of $\exp_+$ in the right-hand side of identity 7 of Suzuki's paper, using successive applications of Eq. (\ref{hadamard}). One gets
\begin{eqnarray}\label{step2}
e^{-s~\mathrm{ad}_{\hB}}\left[e^{-s~\mathrm{ad}_{\hA}}-1\right]\hB&=&e^{-s\hB}\left[e^{-s~\mathrm{ad}_{\hA}}\hB-\hB\right]e^{s\hB}\nonumber\\
&=&e^{-s\hB}\left[e^{-s\hA}~\hB~e^{s\hA}-\hB\right]e^{s\hB}\nonumber\\
&=&e^{-s\hB}~e^{-s\hA}~\hB~e^{s\hA}~e^{s\hB}-\hB.
\end{eqnarray}
%
Let us set
\begin{equation}
\hat{X}(t)=\mathcal{T}\exp_+\left\{\int_0^tds~e^{-s~\mathrm{ad}_{\hB}}\left[e^{-s~\mathrm{ad}_{\hA}}-1\right]\hB\right\},
\end{equation}
and
\begin{equation}
\hat{Y}(t)=e^{-t\hB}~e^{-t\hA}~e^{t(\hA+\hB)}.
\end{equation}
For $t=0$, we have $\hat{X}(0)=\hat{Y}(0)$. In order to prove that $\hat{X}(t)=\hat{Y}(t)\;\;\;\forall~ t$, let us prove the equality of their logarithmic derivative
\begin{equation}\label{loga}
\frac{d\hat{X}(t)}{dt}\hat{X}^{-1}(t)=\frac{d\hat{Y}(t)}{dt}\hat{Y}^{-1}(t).
\end{equation}
We have 
\begin{equation}
\hat{Y}^{-1}(t)=e^{-t(\hA+\hB)}~e^{t\hA}~e^{t\hB},
\end{equation}
and thus
\begin{eqnarray}
\frac{dY(t)}{dt}Y^{-1}(t)&=&\left[-\hB~e^{-t\hB}~e^{-t\hA}~e^{t(\hA+\hB)}-e^{-t\hB}~\hA~e^{-t\hA}~e^{t(\hA+\hB)}\right.\nonumber\\
& &\left.+e^{-t\hB}~e^{-t\hA}(\hA+\hB)~e^{t(\hA+\hB)}\right]\times e^{-t(\hA+\hB)}~e^{t\hA}~e^{t\hB}\nonumber\\
&=&-\hB+e^{-t\hB}~e^{-t\hA}~\hB~e^{t\hA}~e^{t\hB},
\end{eqnarray}
%
%
\begin{eqnarray}
\frac{d\hat{Y}(t)}{dt}\hat{Y}^{-1}(t)&=&-\hB-e^{-t\hB}\hA e^{t\hB}+e^{-t\hB}e^{-t\hA}(\hA+\hB)e^{t\hA}e^{t\hB},\\
&=&-\hB+e^{-t\hB}~e^{-t\hA}~\hB~e^{t\hA}~e^{t\hB},
\end{eqnarray}
and from Eq. (\ref{step2})
\begin{equation}
\frac{\hat{Y}(t)}{dt}\hat{Y}^{-1}(t)=e^{-s~\mathrm{ad}_{\hB}}\left(e^{-s~\mathrm{ad}_{\hA}}-1\right)~\hB.
\end{equation}
Using the first relation (2.3) of Susuki's paper, one has directly
\begin{equation}
\frac{d\hat{X}(t)}{dt}\hat{X}^{-1}(t)=e^{-s~\mathrm{ad}_{\hB}}\left(e^{-s~\mathrm{ad}_{\hA}}-1\right)\hB,
\end{equation}
which proves Eq. (\ref{loga}) and subsequently Eq. (3.45), Identity 7 of Ref. \cite{Suzuki_1985} and Eq. (\ref{eq:integration_zassenhaus}) of the present work.

\end{document}